# On the mechanism of "tulip flame" formation: the effect of ignition sources

Chengeng Qian [a] and Mikhail A. Liberman [b]

[a] *Aviation Key Laboratory of Science and Technology on High Speed and High Reynolds Number, Shenyang Key Laboratory of Computational Fluid Dynamics, Aerodynamic Force Research AVIC Aerodynamics Research Institute,*
*Shenyang 110034, China*

[b] *Nordita, KTH Royal Institute of Technology and Stockholm University, Hannes Alfvéns väg 12, 114 21 Stockholm, Sweden*

Emails:
Chengeng Qian: qiancg@avic.com
Mikhail A. Liberman: mliber@nordita.org

# On the mechanism of "tulip flame" formation: the effect of ignition sources

## Abstract

The early stages of hydrogen-air flame dynamics and the physical mechanism of tulip flame formation were studied using high-resolution numerical simulations to solve the two-dimensional fully compressible Navier-Stokes equations coupled with a one-step chemical model, which was calibrated to obtain the correct the laminar flame velocity- pressure dependence. The formation of tulip flames was investigated for a flame ignited by a spark or by a planar ignition and propagating to the opposite closed or open end. For a flame ignited by a spark on-axis at the closed end of the tube and propagating to the opposite closed or open end, a tulip flame is created by a tulip-shaped axial velocity profile in the unburned gas flow near the flame front caused by the rarefaction wave(s) created by the flame during the deceleration stage(s). It is shown that, in a tube with both closed ends, this mechanism of tulip flame formation also holds for flames initiated by planar ignition. The deceleration stages in the case of planar ignition are caused by collisions of the flame front with pressure waves reflected from the opposite end of the tube. In the case of a flame initiated by planar ignition and propagating toward the open end, the mechanism of tulip flame formation is related to the stretching of the flame skirt edges backward along the side wall of the tube due to wall friction, which leads to the formation of bulges in the flame front near the tube walls. The bulges grow and finally meet at the axis of the tube, forming a tulip-shaped flame. Regardless of the method of flame initiation at the closed end, no distorted tulip flame is formed when the flame propagates to the open end of the tube.

# 1. Introduction

The dynamics of a flame propagating in closed or semi-open tubes, is important for understanding combustion processes under confinement, such as explosions and safety issues as well as for industrial and technological applications, e.g. combustion in gas turbines and internal combustion engines. The inversion of the flame front propagating from the closed end of the tube, from a convex shape with a tip in the direction of unburned gas to a concave shape with a tip in the direction of burned gas, is known as tulip flame formation and has been observed in many experiments and numerical simulations. The first photographs of the inversion of the flame front during the propagation of a premixed flame in a tube were published by Ellis [1, 2, 3] almost a century ago. The concave flame front with lateral petals in the direction of the unburned gas was called the "tulip-flame" by Salamandra et al. [4].

It should be noted that flame front inversion can be caused by various processes, such as the interaction of the flame with a shock or pressure wave, various hydrodynamic instabilities inherent to the propagating flame, such as Darier-Landau (DL) or Rayleigh-Taylor (RT) instabilities. This makes the concept of tulip flame formation not entirely definite, and has led to a number of different scenarios underwent in attempts to explain the mechanism of tulip flame formation. Since the first experimental observation of tulip flames [1-3], various explanations have been suggested in attempts to explain the inversion process leading to the tulip flame, but the actual cause has not been conclusively determined.

Markstein [5, 6] hypothesized that the collision of a convex flame front with a shock wave, leading to an inversion of the flame front, could explain the formation of tulip flames, but the formation of a tulip flame does not involve sufficiently strong shock waves. Many authors [7-18] have considered the DL instability as the cause of flame front inversion, but the characteristic time scale of the DL instability is much larger than the characteristic time scale of the flame front inversion observed in experiments and numerical simulations. Clanet and Searby [19] assumed that the tulip flame formation is a manifestation of the RT instability of the flame front during the flame deceleration. The interaction of vortices in the combustion products near the side walls of the tube, apparently arising due to the baroclinic effect and observed in experiments and modelling, with the flame front was considered by many authors [8, 20--29] as a seemingly plausible scenario of flame front inversion, although no evidence that this is indeed the case was obtained. It has been suggested in [21, 23, 25] that the DL instability acts as a trigger for the appearance of vortices, which then participate in the inversion of the flame front. However, numerical simulations [30, 31] based on an inviscid, zero-Mach number model showed that the tulip flame forms in the absence of vortices. The reader can find a dramatic history of the experimental, theoretical, and numerical studies conducted in an attempt to explain the mechanism of tulip flame formation in the comprehensive review by Dunn-Rankin [32]. It should be noted that, according to more recent experimental studies by Ponizy et al. [33], the formation of a tulip flame is a purely hydrodynamic phenomenon, and the inherent instabilities of the flame front do not participate in the process of the tulip flame formation.

Of particular note is the work of Guénoche [34], who suggested that the rarefaction wave generated by the flame during the deceleration phase is the key process for understanding the mechanism of tulip flame formation. However, with the computational capabilities that existed in the 1960s, it was difficult to prove Guénoche 's conjecture. Only recently, Liberman et al. [35] have shown that the mechanism of a tulip flame formation is indeed close associated with the rarefaction wave arising at the stage of flame deceleration. In [35], the formation of tulip flames in channels with closed ends and different aspect ratios, as well as in a semi-open channel, was investigated by simulation of the fully compressible Navier-Stokes equations combined with a detailed chemical model for a stoichiometric mixture of hydrogen/air. While in the acceleration phase the flame creates a flow of unburned fuel that moves in the same direction ($x > 0$) as the flame, in the deceleration phase the flame front acts like a piston moving out of the tube with negative acceleration and creates a simple rarefaction wave whose head propagates in the direction $x > 0$ at the speed of sound. In the classical case, when the gas ahead of the piston is at rest, the rarefaction wave creates a reverse flow with a maximum (negative) velocity near the piston surface, which vanishes at the head of the rarefaction wave. Therefore, in the case of a flame propagating from the closed end, the flow of unburned gas mixture during the decelerating phase is a superposition of the flow generated earlier by the accelerating flame and the flow created by the rarefaction wave. This leads to a decrease in the velocity of unburned gas flow in the near zone ahead of the flame front and to an increase in the boundary layer thickness. As a result, the axial velocity profile in the unburned flow near the flame front acquires a tulip shape, and since the velocity of each point of the flame front in the thin flame model can be considered as the sum of the laminar flame velocity and the velocity with which the flame is entrained by the unburned mixture flow, the flame acquires a tulip shape.

Usually, the premixed flame is initiated by an electrical spark, which is modeled in simulations as a small circular pocket of hot burned gas near the closed end of a tube. Clanet and Searby [19] identified four stages of flame evolution in the formation of a tulip flame: (1) ignition followed by a hemispherical flame expansion unaffected by side walls of the tube; (2) a finger-shaped flame with exponential growth of the flame surface and, consequently, the flame speed; (3) the lateral parts of the flame skirt touch the side walls of the tube, resulting in a reduction of the flame surface and flame velocity; (4) the flame front inversion and a tulip flame formation.

Obviously, the dynamics of flame propagation in the tube and the evolution of the flame shape depend on how the flame was initiated (ignited). If the flame is initiated by an ignition source whose dimensions are comparable to the width of the tube, the flame skirt touches the tube walls from the very beginning and skips the third stage of deceleration. A flame initiated by a planar ignition source is a representative case opposite to the "classical" ignition from a small spark.

Numerical experiments [36-39] have shown that a tulip-shaped flame is also formed in the case of planar ignition, but the mechanism of tulip flame formation in this case remains unclear. In [39], the formation of cusps on the tulip petals (distorted tulip flame) was explained by "combined effects of the vortex motions and the Rayleigh-Taylor instability driven by pressure waves", but the mechanism and physical processes responsible for the formation of a tulip flame itself from an initially plane flame remained unclear.

The problems discussed below concern the dynamics of premixed flames propagating in the tube with no-slip walls, the effect of the ignition scources, and the mechanism of tulip flame formation in a tube with both closed ends and in a semi-open tube. For this purpose, numerical simulations of the fully compressible reactive Navier-Stokes equations coupled with a one-step chemical model, which was calibrated to correctly describe the pressure dependence of the laminar flame speed in a stoichiometric hydrogen-air mixture, were performed using a high-order numerical code.

## 2. Dynamics of flames in a channel with no-slip walls

We consider a flame which is ignited at the left closed end of the two-dimensional rectangular channel of width $D$ ( $y \in [-D/2, D/2]$ ) and propagates along the x-axis to the right closed or open end. The early stages of flame dynamics ignited by a small spark near the left closed end are well known [19], and the mechanism of tulip-shaped flame formation was for the first time explained in [35] for channels with different aspect ratios and a semi-open channel.

To investigate the dynamics of a flame initiated by planar ignition, consider a flame initiated at the left closed end of the channel by a plane strip of high-temperature combustion products of height $D$ and thickness 1mm. The dynamics of an initially plane flame ignited at the closed end of the channel differs significantly from the dynamics of a flame ignited by a small spark. The dynamics of flame propagating in a tube depends on the boundary conditions and the tube width, which determine the flame acceleration compatible with the boundary conditions. We adopt the adiabatic no-slip boundary condition at the tube walls.

After the flame was ignited near the closed end of the tube, thermal expansion of the high-temperature combustion products pushes the unburned gas towards the opposite closed or open end of the tube, creating a plane-parallel flow of unburned fuel ahead of the flame. At the initial moment, the velocity of the unburned flow is $u = (\Theta - 1)U_f$, and the flame propagates with normal laminar velocity $U_f$ relative to the unburned fuel and with velocity $U_{fL} = \Theta U_f$ in the laboratory reference frame [40, 41], where $\Theta = \rho_u / \rho_b$ is the expansion coefficient – the ratio of the unburned $\rho_u$ and burned $\rho_b$ densities, respectively. Due to the wall friction, the velocity of the unburned flow is maximum near the pipe axis decreases in the thin boundary layer and is zero at the channel walls. The velocity profile in the unburned gas flow depends on the channel width and gas viscosity. In the wide channel, the velocity is uniform in the bulk and drops to zero in the thin boundary layer, while in the thin channel, a parabolic velocity profile (Poiseuille flow) develops in the unburned gas in a short time $t_P \approx D^2 / 100\nu$, so the flame is accelerated all the time and a tulip-shaped flame is not formed in the thin channel [41].

Consider first a flame initiated by planar ignition at the left closed end and propagating to the right open end of the tube. For theoretical analysis, we use the classical approach of an infinitely thin flame front [40, 41]. Every part of the flame front moves relative to the unreacted mixture with a laminar velocity $U_f$ and is simultaneously entrained by the flow of the unreacted mixture with a local flow velocity $u_+(x, y)$ immediately ahead of that part of the flame front. Because of the wall friction, the velocity of the unburned gas mixture vanishes at the channel walls, so the flow field ahead of the flame is not uniform. The shape of the flame front is determined by the relative motion of the various parts of the flame front, which in turn is determined by the velocity profile in the unreacted flow in the near zone ahead of the flame. The local velocity of every small part of the flame front in the laboratory reference frame can be written as[1]

$$\vec{U}_{fL} = \vec{U}_f + \vec{u}_+(x, y). \tag{1}$$

For a wide channel the velocity of unburned flow ahead of the flame $u_+(x, y)$ is nearly uniform in the bulk flow and drops to zero near the side walls within a thin boundary layer of thickness $\delta_l \approx 5X/\sqrt{\text{Re}} \ll D$, where $\text{Re} = u_+(x,0)D/\nu$ is Reynolds number, $\nu$ is the kinematic viscosity of the gas mixture. The shape of the flame front "replicates" the "reversed" shape of the unburned flow velocity profile, remaining nearly flat in the bulk, with the edges of the flame skirt stretched backward within the boundary layer. A stretched flame consumes fresh fuel over a larger area, resulting in a higher rate of heat release per unit of a frontal projected area. The increase in the rate of heat release due to the flame stretching leads to a higher volumetric burning rate, and a higher average heat release rate per frontal area of the flame sheet. It should be emphasized that, unlike a flame ignited by a small spark, the increase in the surface of the flame front initiated by a planar ignition is localized in the narrow strip near the side walls of the tube of width about the boundary layer thickness, which results a higher volumetric burning rate there and in the formation of bulges on the flame front near the tube walls, while the central part of the flame front remains flat (see Fig. 8b). As the velocity is directed along the normal to the flame surface at each point of the flame front, the bulges in the flame front near the tube walls begin to grow. Since the velocity of unburned flow ahead of the flame $u_+(x, y)$ is maximum at $y \geq \delta_l$, the tip of the bulge expands faster in the x-axis direction. The expansion of the bulges towards the tube axis "eats up" the central almost flat part of the flame front until the bulges growing from both sides of the tube axis "eat up" the entire central part of the flame front and converge on the tube axis to form a tulip-shaped flame.

---

[1] Interestingly, although equation (1) is the result of a theoretical model in which the flame front is treated as a discontinuity surface, comparison with numerical simulations shows that this condition is satisfied with a good accuracy (see Appendix B).

In the case of a tube with both ends closed, the evolution of the flame front shape initiated by planar ignition is mostly determined by the collision of the flame with pressure waves reflected from the right closed end of the tube. Each such collision reduces the flame speed, and the decelerating after the collision flame generates a weak rarefaction wave that changes the velocity profile in the unburned flow and thereby favors the formation of a tulip-shaped velocity profile of the unburned gas in the immediate vicinity upstream of the flame front, which in turn causes a tulip profile of the flame front. In addition, the flame collisions with pressure waves leads to distortion of the flame front and the formation of a distorted tulip flame (DTF), which does not occur in a semi-open channel.

## 3. Numerical models

### 3.1 Numerical model

The two-dimensional computational domains that were modelled using high-resolution DNS are $D = 1 \text{cm}$ wide rectangular channels with aspect ratio $L/D = 6$, with both ends closed, and a 1 cm wide rectangular semi-open channel with the right end open. The simulations solve the 2D time-dependent, reactive compressible Navier-Stokes equations including molecular diffusion, thermal conduction, viscosity with a single-step Arrhenius model for chemical reaction of a stoichiometric hydrogen/air mixture. The governing equations are

$$\frac{\partial \rho}{\partial t} + \frac{\partial (\rho u_i)}{\partial x_i} = 0, \tag{2}$$

$$\frac{\partial (\rho u_i)}{\partial t} + \frac{\partial (P\delta_{ij} + \rho u_i^2)}{\partial x_j} = \frac{\partial \tau_{ij}}{\partial x_j}, \tag{3}$$

$$\frac{\partial (\rho E)}{\partial t} + \frac{\partial \left[(\rho E + P)u_i\right]}{\partial x_i} = \frac{\partial (\tau_{ij} u_i)}{\partial x_i} - \frac{\partial q_i}{\partial x_i}, \tag{4}$$

$$\frac{\partial \rho Y_k}{\partial t} + \frac{\partial \rho u_i Y_k}{\partial x_i} = \frac{\partial}{\partial x_i}\left(\rho V_{ik} Y_k\right) + \dot{\omega}_k, \tag{5}$$

$$P = \rho R_B T \left( \sum_{i=1}^{N_s} \frac{Y_i}{W_i} \right), \tag{6}$$

$$\sigma_{xx} = 2\mu \frac{\partial u_x}{\partial x} - \frac{2}{3}\mu\left(\frac{\partial u_x}{\partial x} + \frac{\partial u_y}{\partial y}\right), \tag{7}$$

$$\sigma_{yy} = 2\mu \frac{\partial u_y}{\partial y} - \frac{2}{3}\mu\left(\frac{\partial u_x}{\partial x} + \frac{\partial u_y}{\partial y}\right), \tag{8}$$

$$\sigma_{xy} = \sigma_{yx} = \mu\left(\frac{\partial u_x}{\partial y} + \frac{\partial u_y}{\partial x}\right). \tag{9}$$

The gas mixture is assumed to be an ideal gas, $P = \rho R_B T \left( \sum_{i=1}^{N_s} \frac{Y_i}{W_i} \right)$, with a constant ratio of specific heats $\gamma = c_P / c_V$, and the total energy is

$$E = \frac{P}{\rho(\gamma - 1)} + \frac{1}{2}(u_x^2 + u_y^2) \tag{10}$$

Here $\rho$, $u_i$, $T$, $P$, $E$, $\tau_{ij}$, $q_i$ are density, velocity components, temperature, pressure, total energy, components of viscosity stress tensors, heat flux, $R_B$ is the universal gas constant. $Y_i$, $W_i$, $V_{i,j}$ are the mass fraction, molar mass and diffusion velocity of species $i$. The viscosity coefficients $\mu_k$ are calculated using the standard method [42, 43]. For species $k$

$$\mu_k \left[ \frac{kg}{m \cdot s} \right] = 2.67 \times 10^{-6} \frac{W_k T / 1000}{\sigma_k^2 \Omega^{(2,2)}(T^*)}, \quad (11)$$

where $\sigma_k$ is the collision diameter. The collision integral [42]

$$\Omega^{(2,2)}(T^*) = 1.0313(T^*)^{-0.1193} + (T^* + 0.43628)^{-1.6041} \quad (12)$$

is a function of reduced nondimensional temperature $T^* = \varepsilon T / k_B$. Here $\epsilon$ is the maximum attraction energy between a pair of molecules and $k_B$ is the Boltzmann constant. The viscosity coefficient of mixture is obtained using the semiempirical formula [42, 43]

$$\mu = \frac{1}{2} \left\{ \sum_{k=1}^{N_s} X_k \mu_k + \left( \sum_{k=1}^{N_s} \frac{X_k}{\mu_k} \right)^{-1} \right\}, \quad (13)$$

where $X_k$ is the molar fraction of species $k$.

The coefficient of heat conduction of species $k$ was obtained using [44]

$$\lambda_k = \left( \sum_{i=1}^{4} a_{i,k} \ln(T)^i \right) \cdot \sqrt{T}, \quad (14)$$

with parameters $a_{i,k}$ presented in Table 1.

The coefficient of mixture heat conduction is calculated using the semi-empirical formula similar to Eq. (13). The Lewis number is assumed to be equal to unity for all species, and the species diffusion coefficient is determined as

$$D_k = \frac{\lambda_{mixture}}{\rho c_p}. \quad (15)$$

**Table 1**
Parameters $a_{i,k}$ used in Eq. (14)

|  | $H_2$ | $O_2$ | $H_2O$ | $N_2$ |
|---|---|---|---|---|
| $a_{0,k}$ | -2.34E+00 | 2.58E-01 | -9.83E-01 | 6.33E-03 |
| $a_{1,k}$ | 1.39E+00 | -1.54E-01 | 6.12E-01 | 3.86E-03 |
| $a_{2,k}$ | -3.04E-01 | 3.44E-02 | -1.42E-01 | -2.38E-03 |
| $a_{3,k}$ | 2.93E-02 | -3.36E-03 | 1.44E-02 | 4.00E-04 |
| $a_{4,k}$ | -1.04E-03 | 1.23E-04 | -5.40E-04 | -2.01E-05 |

Recent high-resolution numerical simulations with a detailed chemical scheme for hydrogen-air flames [35] have shown that the formation of tulip flame is a purely hydrodynamic phenomenon. A similar conclusion was obtained earlier by Ponizy et al. [33] as a result of an experimental study of tulip-shape formation in a stoichiometric propane-air flame ignited at the closed end. Therefore, a simplified one-step chemical model that provides a correct pressure dependence of the laminar flame velocity can be used for modelling.

We consider the irreversible global reaction

$$H_2 + 0.5 O_2 \Rightarrow H_2O. \tag{16}$$

The reaction rate is taken in the form a one-step Arrhenius-type chemical kinetics. For pressures $P < 2\,\text{bar}$:

$$\dot{\omega} = dY_{H_2}/dt = A\exp\left(-\frac{E_a}{R_B T}\right)\left(\frac{\rho Y_{H_2}}{W_{H_2}}\right)\left(\frac{\rho Y_{O_2}}{W_{O_2}}\right), \tag{17}$$

while for pressures $P > 2\,\text{bar}$:

$$\dot{\omega} = dY_{H_2}/dt = A\exp\left(-\frac{E_a}{R_B T}\right)\left(\frac{\rho Y_{H_2}}{W_{H_2}}\right)^{0.9}\left(\frac{\rho Y_{O_2}}{W_{O_2}}\right)^{0.9} \tag{18}$$

The reaction order is $n = 2$ for $P < 2\,\text{bar}$, and for $P > 2\,\text{bar}$ the reaction order is $n = 1.8$. A comparison of the laminar flame velocity and flame dynamics obtained from the one-step model Eqs. (17, 18) with the results of the detailed chemical model is given in Appendix A.

### 3.2 Numerical method, boundary conditions, modeling parameters

Two-dimensional (2D) direct numerical simulation (DNS) is used to solve the governing Eqs. (2 - 9) using the DNS solver, which is a weighted essentially non-oscillatory (WENO) fifth order finite difference scheme for solving the convective terms of the governing equations [45]. The advantage of the WENO finite difference method is the capability to achieve arbitrarily high order accuracy in smooth regions while capturing sharp discontinuity. To ensure the conservation of the numerical solutions, the fourth order conservative central difference scheme is used to discretize the non-linear diffusion terms [46]. The time integration is third order strong stability preserving Runge–Kutta method [47].

The initial conditions are $P_0 = 1\,atm$, $T_0 = 298\,K$. In simulations were used adiabatic no-slip reflecting boundary conditions at the tube walls:

$$\vec{u} = 0, \ \partial T / \partial \vec{n} = \partial Y_k / \partial \vec{n} = 0, \tag{19}$$

where $\vec{n}$ is the normal to the wall (y-axis). The nonreflecting outflow boundary condition [48] is employed to calculated subsonic outflow for modelling a semi-open tube.

The reliable simulations of reactive flows require a proper resolution of the inner structure of a flame. Since the thickness of the laminar flame decreases with increasing pressure, higher resolution is required to determine the flame structure at maximum pressure. The maximum pressure during the formation of the tulip flame is about 2 bar in the case of a short tube with both ends closed, which leads to decrease in the thickness of the flame front from $L_f = 350\,\mu m$ to $L_f \simeq 130\,\mu m$. A uniform mesh with resolution $\Delta x \simeq 12.5\,\mu m$, which corresponds to 28 grid points across the flame width at the beginning of the process and 14 grid points at maximum pressure in the case of a tube closed at both ends, was used in the simulations. Thorough resolution and convergence (a grid independence) tests were performed in previous publications [35, 49] by varying the value of $\Delta x$ to ensure that the resolution is adequate to capture details of the problem in question and to avoid computational artefacts. The parameters used in simulations are shown in Table 2.

**Table 2.** Model parameters for simulating a stoichiometric hydrogen–air flame.

| Initial pressure | $P_0$ | 1.0 atm |
|---|---|---|
| Initial temperature | $T_0$ | 298 K |
| Initial density | $\rho_0$ | $8.5 \cdot 10^{-4}$ g/cm$^3$ |
| Pre-exponential factor $P < 2$ bar | $A$ | $2.95 \cdot 10^{13}$ [cm$^3$/mol·s] |
| Pre-exponential factor $P > 2$ bar | $A$ | $2.1 \cdot 10^{12}$ [cm$^{2.4}$/mol$^{0.8}$·s] |
| Activation energy | $E_a$ | $27 R_B T_0$ |
| Laminar flame velocity | $U_f$ | 2.43 m/s |
| Laminar flame thickness | $L_f$ | 0.0325 cm |
| Adiabatic flame temperature | $T_b$ | 2503 K |
| Expansion coefficient ($\rho_u / \rho_b$) | $\Theta$ | 8.34 |
| Specific heat ratio | $\gamma = C_P / C_V$ | 1.399 |
| Sound speed | $a_s$ | 408.77 m/s |

## 4. Results of simulations

### 4.1 Tulip flame formation in 2D rectangular channel with both ends closed

We consider the tulip flame formation in the two-dimensional rectangular channel of width $D = 1 cm$, aspect ratio $L/D = 6$ with both ends closed. Fig. 1a and Fig. 1b show the time evolution of the local velocities of the flame front along the center line $U_{fL}(y=0)$ and near the side wall for the case of a flame ignited by a small spark at the tube axis at the left closed end (Fig.1a) and for the flame ignited by a planar strip of high temperature combustion products at the left closed end (Fig.1b).

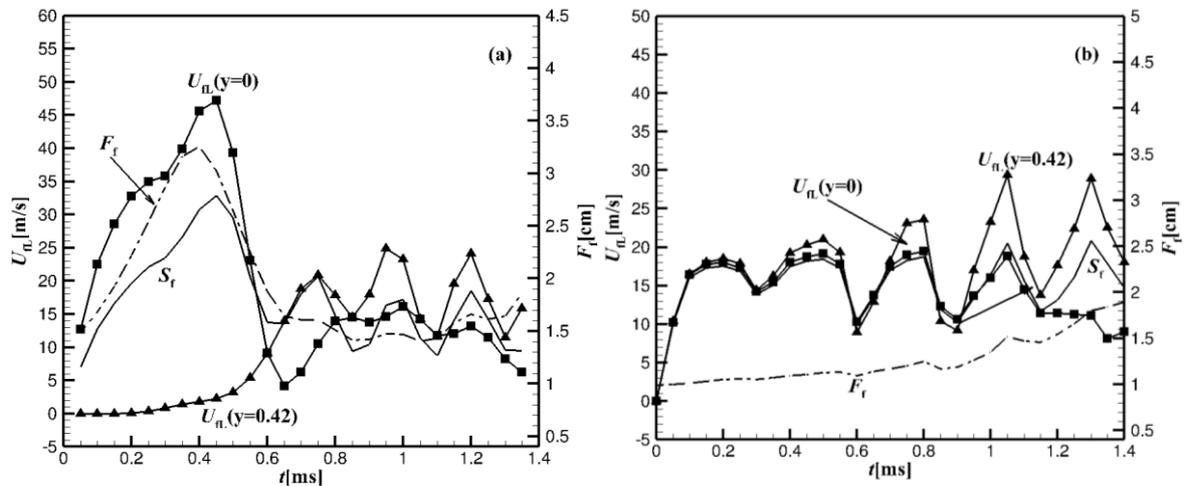

**Fig. 1a**. Time evolution of the flame surface area $F_f$ (dashed-dotted), combustion wave speed $S_f$, local velocities of the flame front at the tube axis $y=0$ and at near the wall $y=0.42cm$ for a flame initiated by a small spark at the axis at the left end of the tube.

**Fig. 1b**. The same as in Fig. 1a but for the flame initiated by planar ignition.

It can be seen in Fig. 1a that the flame propagation speed in the laboratory reference frame is closely related to the variation of the flame surface area $F_f$. From the beginning and up to 0.5ms the flame surface area $F_f$, the combustion wave velocity $S_f$ as well as the local velocities of the flame front $U_{fL}(y=0)$ and $U_{fL}(y=0.42\text{cm})$ increase. After the lateral parts the flame skirt touch the side walls of the tube, the flame surface area $F_f$, combustion wave velocity $S_f$ and flame tip velocity $U_{fL}(y=0)$ decrease, however the local flame front velocity near the wall at $y=0.42\text{cm}$ continues to increase and after 0.65ms it exceeds the flame tip velocity, $U_{fL}(y=0)$.

The dynamics of the flame shown in Fig. 1b, which is initiated by planar ignition, is quite different. The flame surface area $F_f$ increases with small oscillations, while the combustion wave velocity $S_f$, the flame tip velocity $U_{fL}(y=0)$, and the local flame front velocity near the wall, $U_{fL}(y=0.42cm)$ strongly oscillate as the result of the flame collisions with pressure waves reflected from the left closed end of the tube, similar to oscillations in Fig. 1a after 0.7ms. Nevertheless, it is seen in Fig. 1b, that after each oscillation the local flame front velocity near the wall, $U_{fL}(y=0.42cm)$ increases, while the flame tip velocity, $U_{fL}(y=0)$ oscilates remaining in avaredge constant and later decreases. Finally, after several collisions after 1.1ms, the velocity $U_{fL}(y=0.42cm)$ significantly exceeds the flame tip velocity $U_{fL}(y=0)$.

Figures 2a and 2b show the evolution of the axial flow velocities of the unburned fuel at 0.5mm ahead of the flame front at the tube axis $y=0$ and near the side wall of the tube, at $y = 0.42\,\text{cm}$ for the case of a spark ignition (Fig. 2a) and for the case of a planar ignition (Fig.2b). In both figures 2a and 2b also shown the difference between the local velocity of the flame front near the side wall of the tube and at the tube axis. Fig. 2a shows that after 0.6ms, $\Delta u_+ = u_+(y = 0.42) - u_+(y = 0)$ becomes positive, and from this moment the parts of the flame front located closer to the tube walls are entrained by the flow of unreacted mixture with higher speed, and, accordingly, the parts of the flame front located closer to the tube walls propagate faster than the parts of the flame front located closer to the tube axis.

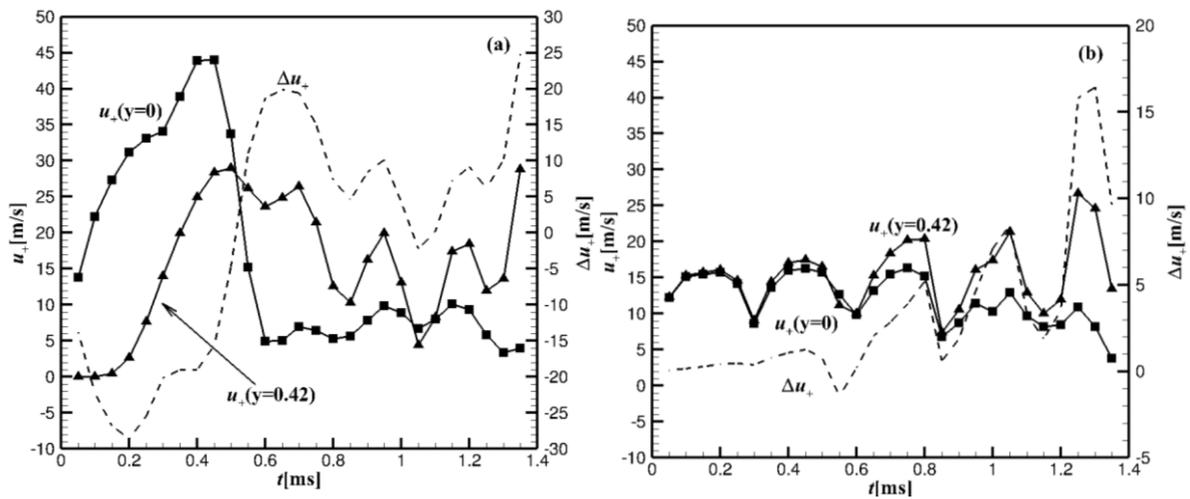

**Fig. 2a**. Velocities of the unreacted flow at 0.5 mm ahead of the flame front at $y = 0$; 0.42 mm, and the difference $\Delta u_+ = u_+(y = 0.42) - u_+(y = 0)$ for spark ignition.
**Fig. 2b**. Velocities of the unreacted flow at 0.5 mm ahead of the flame front at $y = 0$; 0.42 mm, and the difference $\Delta u_+ = u_+(y = 0.42) - u_+(y = 0)$ for planar ignition.

It was shown in [35] that the difference $\Delta u_+ = u_+(y = 0.42) - u_+(y = 0)$ between the axial velocities in the unburned flow mainly arises due to the first strongest rarefaction wave. When the flame surface decreases, the combustion wave velocity also decreases. The decelerating flame acts like a piston moving with (negative) acceleration out of a tube and creates a simple rarefaction wave in the unburned gas mixture. In Fig. 1a one can see oscillations of the flame velocity caused by collisions of the flame front with the pressure waves reflected from the right end of the tube. Similar oscillations in the velocity of unburned gas near the flame front, caused by the collision of the flame with pressure waves reflected from the right end of the tube, can be seen in Fig. 2a and Fig. 2b. As a result of these collisions, the flame speed decreases, causing the flame to decelerate, which creates a weak rarefaction wave in the unburned gas, while the pressure wave reflected from the left end of the tube accelerates the flame. One of the consequences of the rarefaction waves in the unburned gas is the adverse pressure gradient in the unburned flow near the flame front. The pressure waves reflected from the right end of the tube, creates conditions similar to those created by the first rarefaction wave described above, as a result of the reduction of the flame surface area (Fig. 1a). The effect of the rarefaction wave produced by each collision of the flame and pressure wave is weaker compared to the first strong rarefaction wave, but repeated collisions strengthen the overall effect, so that the value of $\Delta u_+ = u_+(y = 0.42) - u_+(y = 0)$ oscillates, but remains positive in both cases shown in Fig. 2a and Fig. 2b already after the formation of the first rarefaction wave. This leads to the formation of a tulip-shaped axial velocity profile in the unburned gas near the flame front, which in turn leads to the formation of a tulip-shaped flame.

In the case of a flame initiated by planar ignition, there is no stage of strong reduction of the flame surface (see Fig. 1b) and corresponding strong flame deceleration, but several collisions of the flame front with pressure waves reflected from the right closed end of the tube are enough to create conditions for the unburned gas velocity near the tube wall $u_+(y = 0.42)$ to increase in the close vicinity ahead of the flame front more than the unburned gas velocity at the tube axis $u_+(y = 0)$. Fig. 2b shows that the value of $\Delta u_+ = u_+(y = 0.42) - u_+(y = 0)$ became positive already after the first collision, fluctuates but grows and becomes positive after 0.85 ms. After 1.2 ms, the value of $\Delta u_+$ became large enough, and the axial velocity profile in the unburned gas close ahead of the flame front became tulip-shaped, and, consequently, the flame front also acquired a tulip-shape.

Figs. 3a and 3b show the calculated Schlieren images for selected time moments and the stream lines during the formation of a tulip flame for the flame propagating in the tube $D = 1\,cm$, $L = 6\,cm$ with both ends closed for spark ignition (Fig. 3a) and for planar ignition (Fig. 3b).

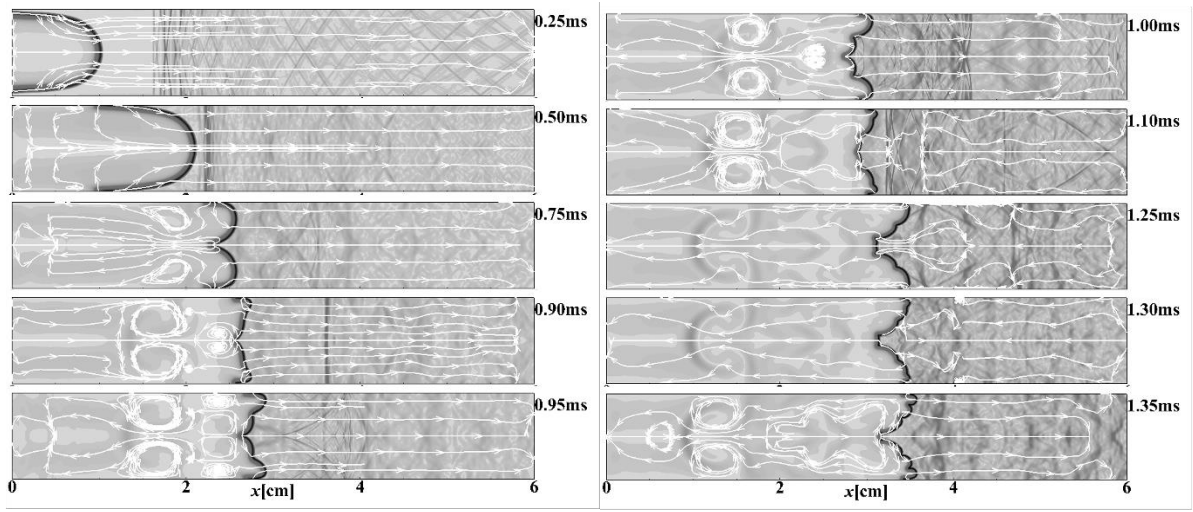

**Fig. 3a**. Time sequence of computed Schlieren images and streamlines for the spark ignited flame propagating in the tube $L/D = 6$ with both ends closed.

In Fig. 3a it is seen that after 0.5 ms, when the flame surface area begins to decrease and thus the flame velocity decreases, the flow velocities in the combustion products are reversed from positive to negative. The reverse flow of combustion products is necessary to fulfill the boundary conditions on the flame surface with a decrease in the flame speed. The reverse flow also causes the formation a pair of vortices in combustion products.

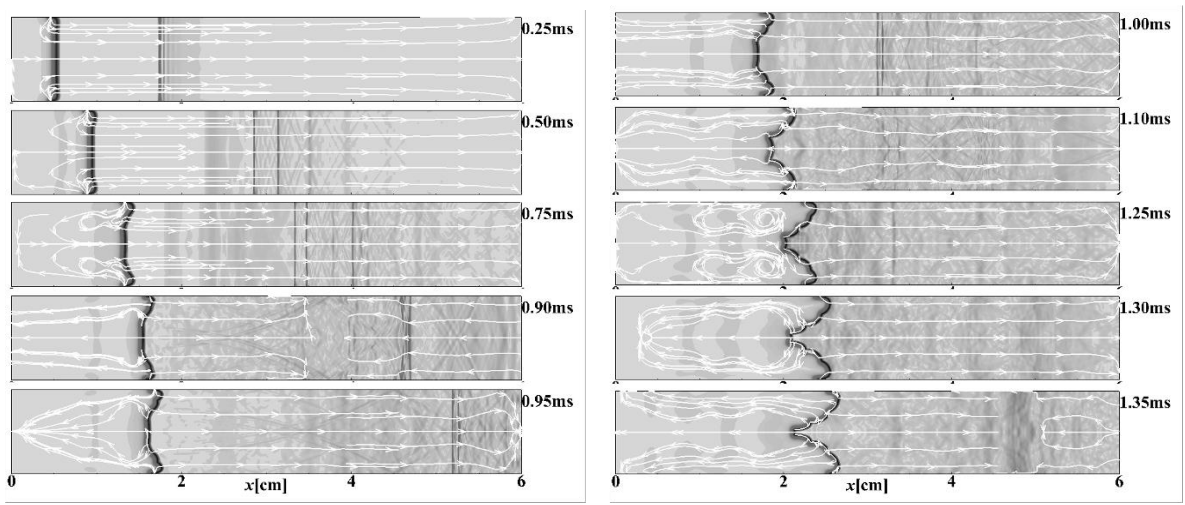

**Fig. 3b**. Time sequence of computed Schlieren images and streamlines for the planar ignition of the flame propagating in the tube $L=6cm$ with both ends closed.

As can be seen in Fig. 3a and Fig. 3b, in both cases of the flame ignited by a small spark for the flame initiated by planar ignition, the pressure waves reflected from the right closed end of the tube and colliding with the flame front cause a significant deformation of the flame leading to wrinkling of the tulip petal surface and formation of distorted tulip flames [29, 39].

It can also be seen in Fig. 3a and Fig. 3b that a pair of vortices created by strongly curved sections of the flame front near the side walls, apparently due to the baroclinic effect, appears in combustion products near the tube walls behind the flame front. But while vortices in Fig. 3a appear at 0.75ms, and remain up to 1.2ms, just before the tulip flame formation, in Fig. 3b they appear at 1.25 ms, when the reverse flow in combustion products already existed as it is seen in slides 0.75, 0.9, 0.95, 1.0, and 1.1 ms. These vortices disappear during the final phases of the tulip flame formation. Many authors [12, 20, 21, 25, 26, 29] believed that vortices play an important role in the formation of a tulip flame. However, the present study convincingly shows that vortices are not relevant to the mechanism of tulip flame formation. In the case of spark ignition (Fig.3a) the boundary conditions require the occurrence of recirculation in the reverse flow of combustion products near the left closed end of the channel. These large-scale vortices subsequently drift in the combustion products toward the flame front, forming a focus of the flow velocity streamlines near the axis of the tube, that apparently extend in both directions from the flame front, as can be seen in Fig. 3a at 0.90 ms and later. A similar scenario is involved in the appearance of large-scale vortices in the case of planar ignition in Fig. 3b. Obviously, this is a purely hydrodynamic process, which does not involve the instabilities of the flame front. A more detailed discussion of this problem is given in [35].

Fig. 4a and 4b show the axial velocity profiles in the flow of unburned gas close ahead of the flame front during the tulip flame formation.

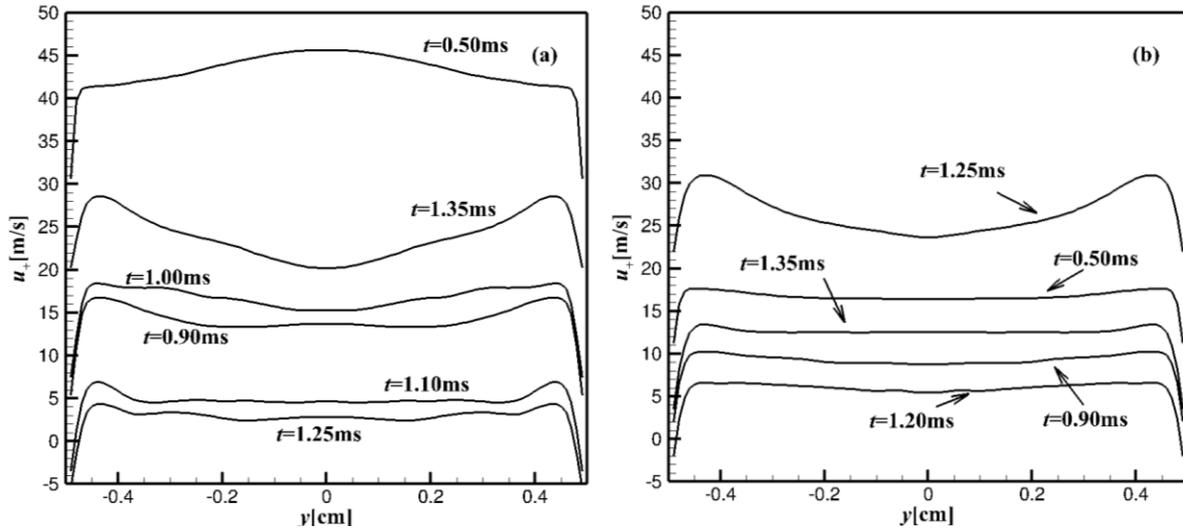

**Fig. 4a**. Profiles of the axial velocities in the unburned gas mixture ahead of the flame front during the tulip flame formation for the flame initiated by a small spark.

**Fig. 4b**. The same as in Fig. 4a, but for the flame initiated by planar ignition.

In [39] the influence of the tube size on flame dynamics was investigated by simulations of a flame initiated by a planar ignition in 2D channels of different sizes $2 \times 14\,cm^2$, $4 \times 28\,cm^2$, $8 \times 56\,cm^2$ and the same aspect radio $L/D = 7$. It was found that the flame dynamics and the normalized characteristics of the flame evolution are essentially the same. This statement is obvious and to some extent misleading because important is not a size of the channel but the aspect radio. For example, it is clear that for the closed tubes of the same width but different lengths, there will be more collisions of the flame with reflected pressure waves in a shorter tube than in a longer tube. To illustrate this, Fig. 5 shows dynamics of the flame (the flame tip velocity at the center line $y = 0$) in the two-dimensional channel computed in [35] for the flames propagating in tubes with both closed ends of the same width but various aspect ratios $L/D = 6, 12, 18$ and for the flame in the half-open tube.

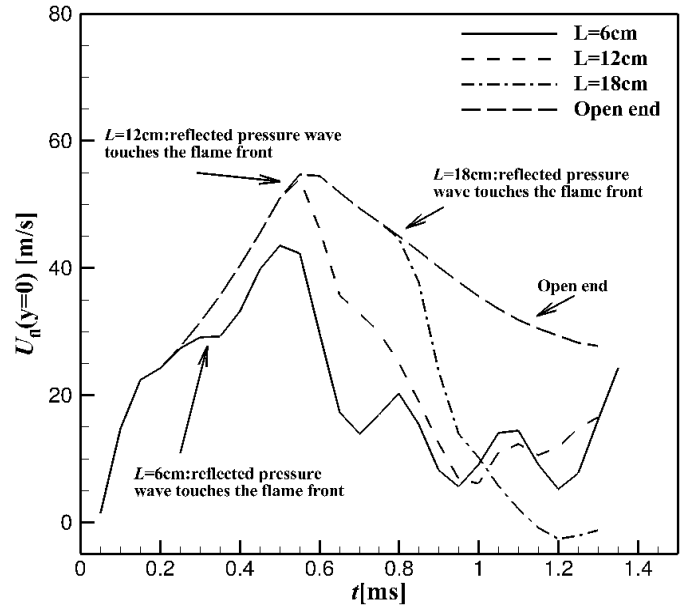

**Fig. 5.** The velocity of the flame tip along the center line $y=0$, computed for tubes with different aspect ratios $L/D = 6, 12, 18$ and for the half-open tube of the same width.

### 4.2 Tulip flame formation in semi-open tubes

In the case of a semi-open tube the flame ignited at the left closed end propagates to the right open end, therefore there are no reflected pressure waves, and the pressure ahead of the flame remains almost constant. Figures 6a and 6b show the time evolution of the flame surface area $F_f$, the speed of the combustion wave $S_f$, and the local velocities of the flame front at the central line and near the sidewall, at $y = 0.40\,cm$ for spark ignition (Fig. 6a) and $y = 0.42\,cm$ for planar ignition (Fig. 6b).

Fig. 6a shows that the acceleration stage of the spark-initiated flame in a semi-open channel is almost the same as for the spark-initiated flame in a channel with both closed ends in Fig. 1a. However, the deceleration rate in the case of a half-open tube is noticeably slower than in a tube with closed ends, the flame deceleration stage in a half-open tube lasts almost twice as long as that in a tube with both closed ends. This means that the intensity of the rarefaction wave created by the flame during deceleration is weaker. In addition, in the case of a semi-open tube, there are no reflected pressure waves, which in a tube with closed ends enhance the effect of the first rarefaction wave. Thus, in a half-open tube only the first rarefaction wave affects the dynamics of the flow ahead of the flame and, consequently, the formation of the tulip flame.

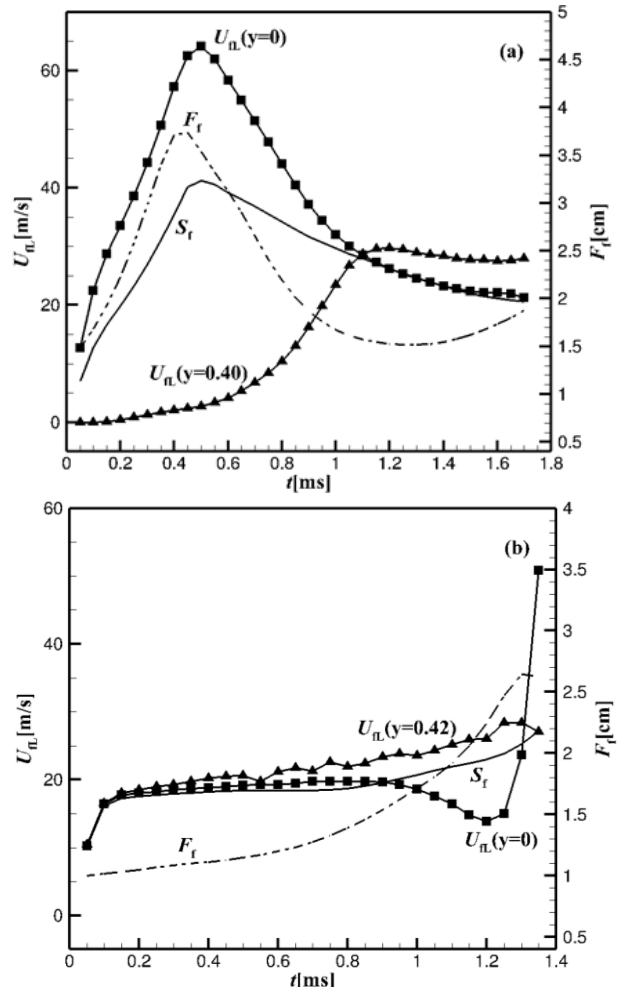

**Fig. 6a**. Calculated time evolution of combustion wave velocity $S_f$, the flame surface area $F_f$ (dashed-dotted) and local velocities of the flame front at $y=0$ and near the sidewall at $y=0.4\,\text{cm}$ in the half-open tube for the flame initiated by a spark.
**Fig. 6b**. The same as in Fig. 6a but for the flame initiated by planar ignition.

In the case of a flame initiated by planar ignition at the closed end and propagating towards the open end (Fig. 6b), there are also no reflected pressure waves and no oscillations of the flame velocity. In this case, there is no deceleration stage, the flame surface area increases monotonically until the tulip flame is formed. Therefore, in the case of a planar flame ignition in a semi-open tube, the mechanism of tulip flame formation is different from that of a spark-ignited tulip flame. In the case of planar ignition in a semi-open tube, the lateral parts of the flame front are stretched backwards along the boundary layer, while the central part of the flame front remains flat (Fig. 8b). The stretched part of the flame front near the wall consumes fresh fuel over a larger area in the strip $\Delta y \approx \delta_l$ near the wall, which leads to an increase in the heat release rate per unit area of the frontal projection. An increase in the heat release rate leads to an increase in the volumetric combustion rate and the formation of bulges on the flame front near the tube walls. It is obvious that the tip of the bulge propagates in the x-axis direction faster than the flat central part of the flame, and the velocity of the unburned flow ahead of the flame is also maximum at $y \approx \pm\delta_l$. The expansion of the two bulges in the direction of the tube axis "eats up" the central almost flat part of the flame front until the bulges growing on both sides of the tube axis "eat up" the entire central part of the flame front and converge on the tube axis to form a tulip-shaped flame.

Figs. 7a and 7b show the flow velocities along the tube axis $y = 0.0$ and near the wall at $y = 0.40 cm$ and $y = 0.42 cm$ in the unreacted gas at 0.5 mm ahead of the flame front for the spark ignited (7a) and planar ignited (7b) flames propagating in the semi-open tubes.

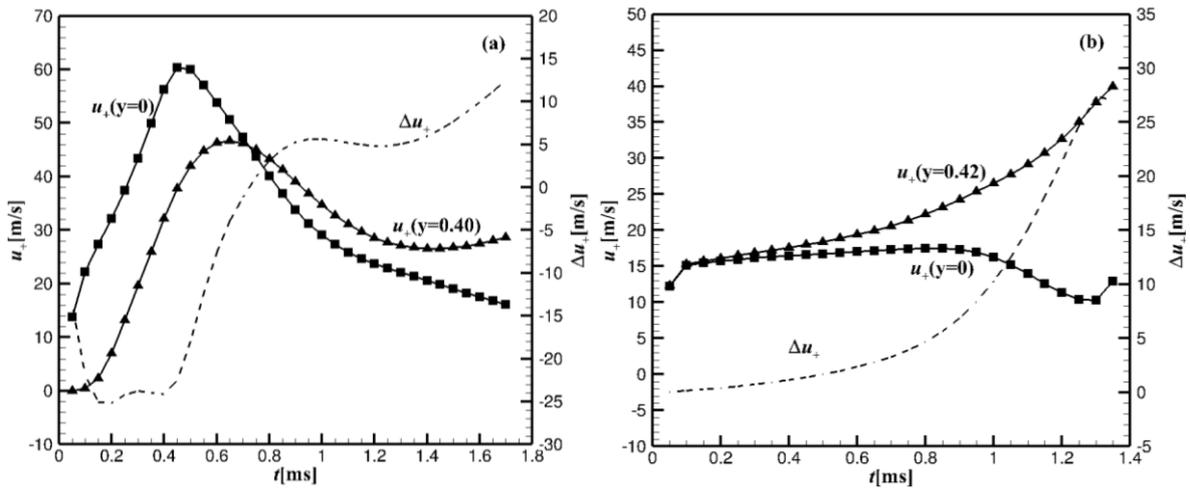

**Fig. 7a**. Velocities of the flow at 0.5 mm ahead of the flame front at $y = 0.0$ and at $y = 0.40\,\text{cm}$, and the difference $\Delta u_+ = u_+(y = 0.40) - u_+(y = 0)$ for the case of flame initiated by the spark ignition.
**Fig. 7b**. Velocities of the flow at 1 mm ahead of the flame front at $y = 0.0$ and at $y = 0.42\,\text{cm}$, and the difference $\Delta u_+ = u_+(y = 0.42) - u_+(y = 0)$ for the case of flame initiated by planar ignition.

Figures 8a and 8b show the sequences of the calculated Schlieren images for selected time moments and the streamlines during the formation of a tulip flame for the flame propagating in the semi-open tubes $D = 1\,cm$ for the case of spark ignition (Fig. 8a) and for planar ignition (Fig. 8b).

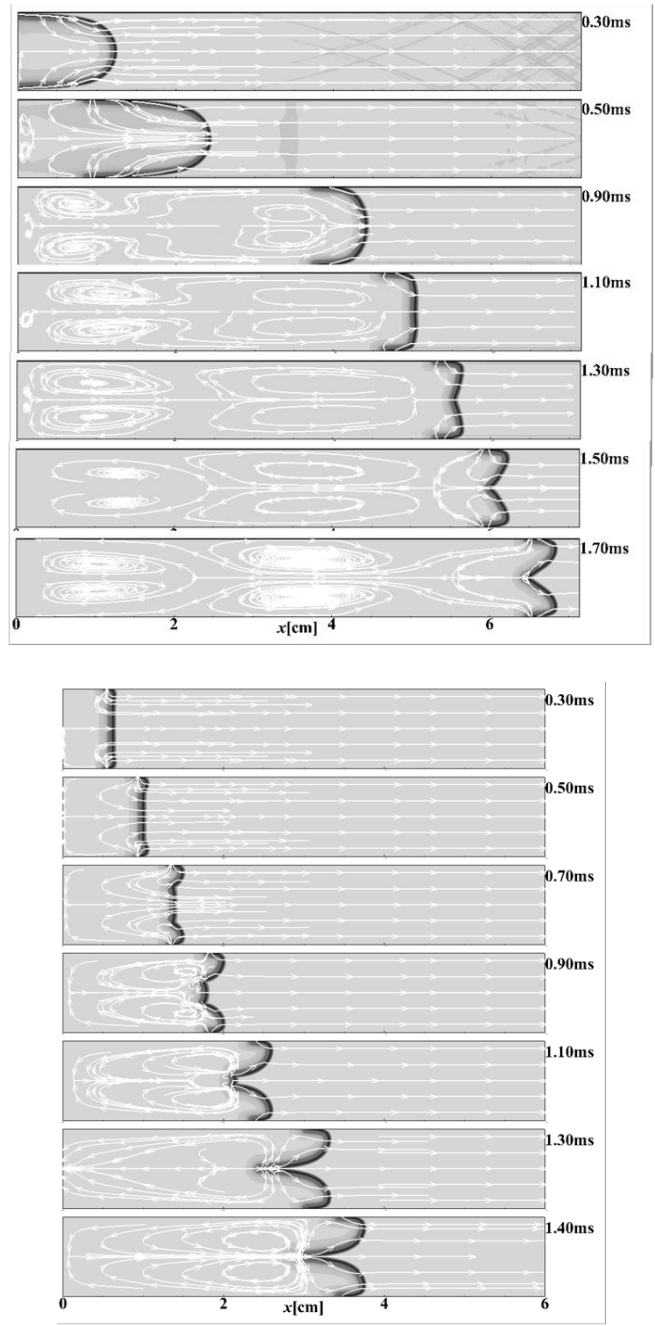

**Fig. 8a**. Sequences of computed schlieren images and the streamlines for the premixed hydrogen–air flame propagating in a half-open tube for the flame initiated by the spark ignition.

**Fig. 8b.** Sequences of computed schlieren images and the streamlines for the premixed hydrogen–air flame propagating in a half-open tube for the flame initiated by the planar ignition.

Fig. 9 shows the axial velocity profiles in the flow of unburned gas ahead of the flame front during the tulip flame formation for the flame initiated by spark ignition. It can be seen that after 1ms, the flow velocities in the unburned gas closer to the side walls exceed the flow velocity at the center line, and the further away from the center line the greater the difference $\Delta u_+$. This trend continues all the way up to the "inner" edge of the boundary layer, where the flow velocity closer decreases inside the boundary layer and vanishes at the side wall of the channel. Finally, the profile of the axial velocity in the unburned mixture ahead of the flame takes the form in inverted tulip. Accordingly, the local velocities of the flame front $U_{fl} = U_f + u_+$ becomes minimum at the central line, gradually increases toward the boundary layer where it becomes maximum and decreases inside the boundary layer, so the flame front also acquires a tulip shape.

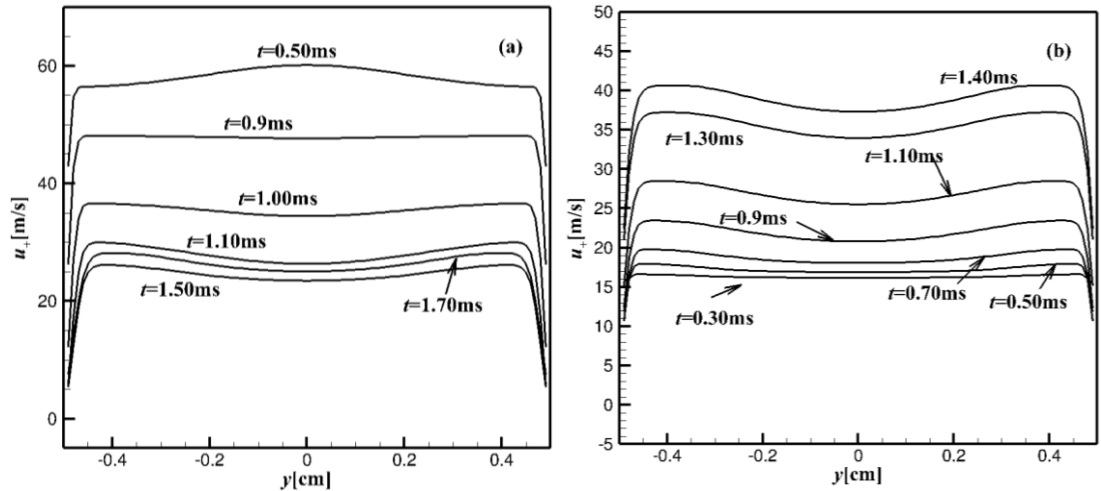

**Fig. 9a.** Axial velocity profiles in the unburned gas for spark ignition.
**Fig. 9b.** Axial velocity profiles in the unburned gas for planar ignition.

## 5. Discussion and conclusions

This paper presents numerical modelling of the early stages of hydrogen-air flame dynamics and a study of the physical processes that are the mechanism of tulip-shaped flame formation for flames propagating in tubes with both ends closed and in semi-open tubes. The two-dimensional fully compressible reactive Navier-Stokes equations combined with a one-step

Arrhenius chemical model, calibrated to obtain the correct pressure dependence of laminar flame velocity, and the ideal gas equation of state were solved using a high-resolution finite-difference (WENO) scheme to capture important features of flow and reaction waves. One of the main purposes of the study was to investigate the influence of the ignition geometry on the mechanism of tulip flame formation. It was considered the formation of the tulip flame ignited by a small spark near the closed (left) end and propagating to the opposite closed or open end and compared with the formation of the tulip flame initially initiated by a planar ignition region at the closed end and propagating to the opposite closed or open end of the tube. The simulations have shown that the early stages of the flame dynamics modelled using a one-step Arrhenius chemical model with a correct pressure dependence of the laminar flame velocity do not differ practically from the flame dynamics obtained by simulations using a detailed chemical model [35].

It was shown that for the flames initiated by a spark ignition at the closed end and propagating in a tube with both ends closed as well as propagating to the open end of the tube, the flame front inversion and the tulip flame formation is exclusively due to the generation of rarefaction wave by the flame front at the stage of flame deceleration. The decelerating flame acts like a piston moving out of a tube with negative acceleration and creates a simple rarefaction wave that creates a reverse (negative) flow of the unburned gas with maximum (negative) velocity near the flame front. The result of superposition of the forward flow of unburned gas created by the flame during the ignition and acceleration stages and the flow created by the rarefaction wave leads to a decrease in velocity in the unburned flow and an increase in the boundary layer thickness. The most important consequence of this is that the superposition of the pre-existing forward flow of the unburned gas with the reverse flow created by the rarefaction wave leads to the formation of an inverse tulip-shaped profile of the unburned gas axial velocity near the flame front. In the theoretical model of the infinitely thin flame front the velocity of any point on the flame front is the sum of the laminar flame speed relative to the unburned gas $\vec{U}_f$ plus the velocity $\vec{u}_+(\vec{r})$ with which this part of the flame front is entrained by the unburned flow $\vec{U}_{fL} = \vec{U}_f + \vec{u}_+$. This means that when the axial velocity profile of the unburned gas close ahead of the flame front takes the tulip shape, the flame front takes a tulip shape also. It should be noted that although the condition $\vec{U}_{fL} = \vec{U}_f + \vec{u}_+$ is valid for the theoretical model of an infinitely thin flame, and in simulations we consider a flame front of real thickness, this condition is nevertheless satisfied with good accuracy (see Appendix B).

For a flame initiated by planar ignition, the tulip flame formation mechanism for a flame propagating in a tube with both ends closed is similar to that described above. In this case, pressure waves reflecting off the right end of the tube and colliding with the flame cause several short stages of flame deceleration. For a spark-initiated flame in a tube with both ends closed the collision of the flame with reflected pressure waves enhance the effect of the first rarefaction wave, which is produced when the flame skirt touches the side walls of the tube, which leads to the decrease of the flame surface area. It should be emphasized that for a spark ignited flame in a half-open tube, only the first rarefaction wave creates a condition for the flame front inversion.

The physics of tulip flame formation is different in the case where the flame is initiated by planar ignition and propagates towards the open end of the tube. In this case, the thermal expansion of the combustion products behind the initially flat flame front creates a flow of unburned gas towards the open end of the tube, where the unburned gas flows out of the tube. Therefore, in this case there are no pressure waves reflected from the opposite end of the tube. The lateral sides of the flame front touch the walls tube from the very beginning and the skirt of the flame front starts to stretch backward within the boundary layer. The increase in the rate of heat release due to the flame stretching leads to a higher average heat release rate per frontal area of the flame sheet in the narrow strip near the side walls of the tube of width about the boundary layer thickness, which results in the formation of bulges on the flame front near the tube walls, while the central part of the flame front remains flat (see Fig. 8b). The velocity of the bulge tips is greater than the velocity of the flame front near the tube axis, where the flame front remains nearly flat. The growing bulges expand from both sides towards the pipe axis, meet at the tube axis and form a tulip-shaped flame.

In the calculated Schlieren images Fig. 3a and Fig. 3b one can see the cusps formed on the tulip petals, which is called a distorted tulip flame (DTF). Obviously, the formation of a distorted tulip flame is due to collisions of the flame with pressure waves reflected from the opposite closed end of the tube. It should be emphasized that no DTF occurs in the semi-open channel (Fig.8 a, b). In [29, 39], the formation of ledges on the tulip petals and the distorted tulip flame are explained by "*the joint action of vortex motions and Rayleigh-Taylor (RT) instability due to the pressure wave*". Figs. 10a and 10b show the flame acceleration in the tube with both ends closed for a spark ignited flame (Fig. 10a) and for the planar ignited flame (Fig. 10b).

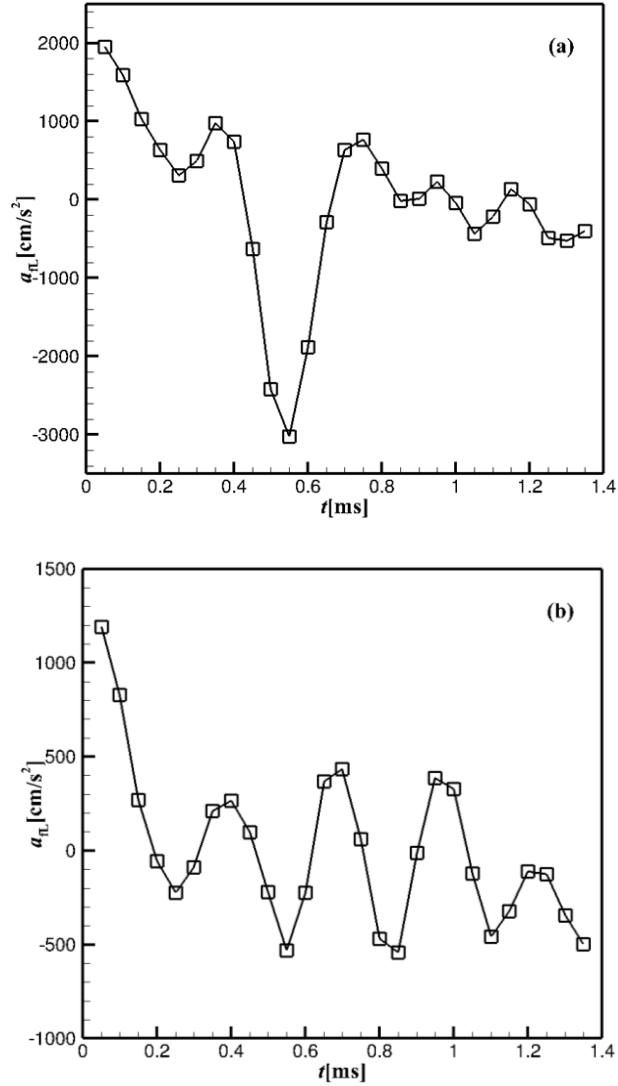

**Fig. 10a**. Spark ignited flame acceleration during a distorted tulip flame formation in the tube with both ends closed.

**Fig. 10b.** Planar ignited flame acceleration during a distorted tulip flame formation in the tube with both ends closed.

The increment of the linear (initial) stage of RT instability is $\sigma = \sqrt{Agk}$, where $A = \dfrac{(\rho_u - \rho_b)}{(\rho_u + \rho_b)} = \dfrac{(\Theta - 1)}{(\Theta + 1)} \approx 0.77$ is the Atwood number, g is the flame acceleration, $k = 2\pi / \lambda$ is the wave number [41]. Taking a typical wave length as the minimum size of the cusp in Figs. 3a, b, $\lambda \approx 1mm$, we obtain $\sigma \simeq 3.48 \cdot 10^2 s^{-1}$ for spark ignited flame (Fig. 10a), and $\sigma \simeq 1.55 \cdot 10^3 s^{-1}$ for planar ignited flame (Fig. 10b). In both cases, the acceleration that can lead to the development of RT instability lasts about $\Delta t \sim 0.1 ms$. Therefore, the amplitude of the RT instability does not grow in both cases, $\exp(\sigma_\Delta t) \approx 1$. However, collisions of the flame with pressure waves are a probable cause of the formation of distorted tulip flames, although this process is not completely clear at present.

Another misleading statement often found in the literature is that the vortices generated behind the flame skirt during flame deceleration, "*expand with time and overtake the flame front, creating conditions that allow the flame to invert*" [39]. The vortices behind the flame front and the inverse flow in the combustion products are a consequence of the boundary conditions for the flow of combustion products and are not related to the inversion of the tulip flame. While the reverse flow in the unburned gas in the near region of the flame front, as well as the adverse pressure gradient, are natural characteristics of the rarefaction waves generated by the flame during the deceleration phases and are the main physical process leading to the formation of the tulip flame.

Our conclusion is that the physical process causing flame front inversion and tulip-shaped flame formation is the formation of a tulip-shaped profile in the unburned gas flow in the near region of the flame front under the action of the rarefaction wave generated by the flame at the flame braking stage. This scenario works in the case of a spark ignited flame in a tube with closed ends and in a semi-open tube, as well as in the case of a planar ignited flame in a tube with closed ends, where the flame deceleration results from the collision of the flame with pressure waves reflected from the opposite end of the tube. In the case of a flame initiated by planar ignition, there is no deceleration stage associated with the reduction of the flame surface due to the extinguishing of the lateral parts of the flame front touching the tube walls. However, when the flame front collides with the pressure wave reflected from the opposite end of the tube, a number of decelerating stages occur. A special mechanism of tulip flame formation occurs when the flame is initiated by planar ignition and propagates towards the open end of the tube. In the later case, the flame front is stretched along the side walls of the tube and the flame accelerates until a tulip flame is formed. In this case, the "tulip" shape is formed by the growth and convergence of two bulges that are formed at the flame front near the tube walls.

## Acknowledgment

No specific funds were received for this particular work; research of Nordic Institute for Theoretical Physics (NORDITA) is partially supported by Nordforsk. (M.L.)

## Disclosure statement

No potential conflict of interest was reported by the authors.

## Appendix A. 1-step chemical model

The one-step chemical model was calibrated to reproduce correctly the most important parameters of a hydrogen/air flame, such as the laminar flame velocity and thickness, the pressure dependence of the laminar flame speed, adiabatic flame temperature, expansion coefficient and sound speed. Fig. A1 shows the laminar velocity of a hydrogen air flame calculated using the present one-step model compared with the detailed chemical model [50]. The pre-exponential constant $A$ and activation energy $E_a$ are calibrated by genetic algorithm method using Cantera and are shown in Table 2. During the formation of a tulip flame, the pressure increase is negligibly small in the case of a semi-open tube, but it increases from an initial value $P_0 = 1\,atm$ up to and above $\approx 2.0\,atm$ in a short channel $L = 6\,cm$.

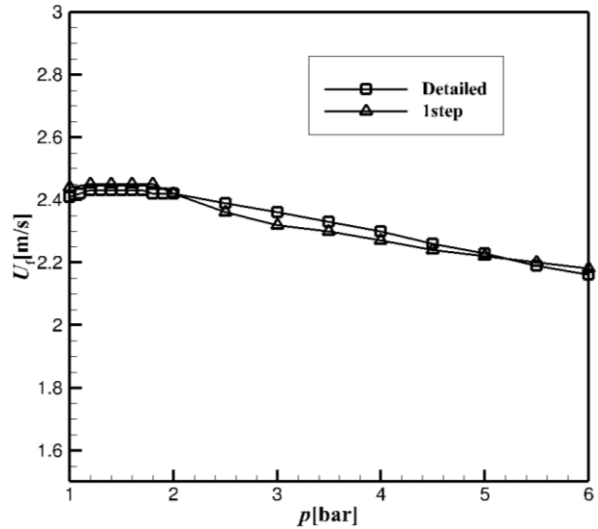

**Fig. A1**. Laminar velocity of a hydrogen air flame vs pressure calculated using a one-step model, Eqs. (17, 18) and detailed chemical model [50].

Comparison of the flame dynamics obtained in simulation with a detailed chemical model [50] and with a one-step chemical used in this paper is shown in Fig. A2.

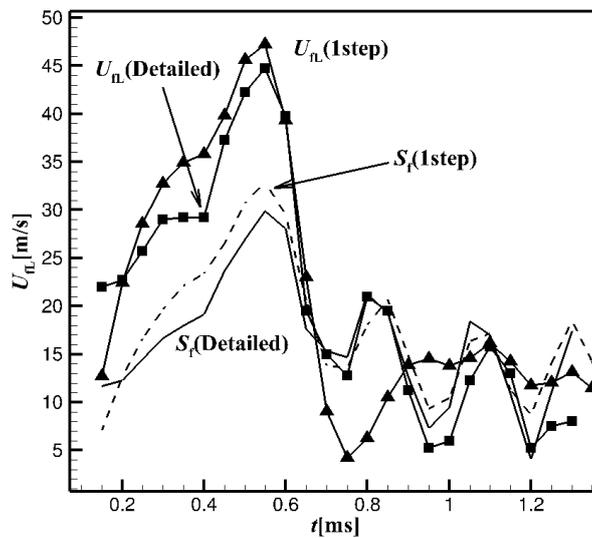

**Fig. A2.** Flame velocity calculated using detailed chemical model [50] and the one-step model Eqs. (17, 18).

It can be seen from Fig. A2 that the flame dynamics, temporal evolution of flame velocity, flame surface area, etc., obtained from simulations with the one-step model are almost the same as those obtained from simulations with the detailed chemical model [50].

## Appendix B. Verification of a thin flame front model

As it was emphasized earlier, in the theoretical model of an infinitely thin flame, the velocity of every point $(x, y)$ on the flame front in the laboratory reference frame can be considered as the sum of the laminar flame speed $U_f$ relative to the unburned gas plus the flow velocity $u_+(x, y)$ immediately ahead of that point of the flame front, with which this part of the flame front is entrained by the flow of the unreacted mixture. Fig. B1 shows the flame front velocity at the axis of a half-open tube for a planar ignited flame (see Fig. 6b), and the unburned gas velocity on the axis 0.5mm ahead of the flame front, $u_+(X_f + 0.5mm, y = 0)$. It is seen that the difference $U_{fL}(X_f, y = 0) - u_+(x = X_f + 0.5mm, 0)$, shown by the empty squares, is quite close to $U_f = 2.41 m/s$.

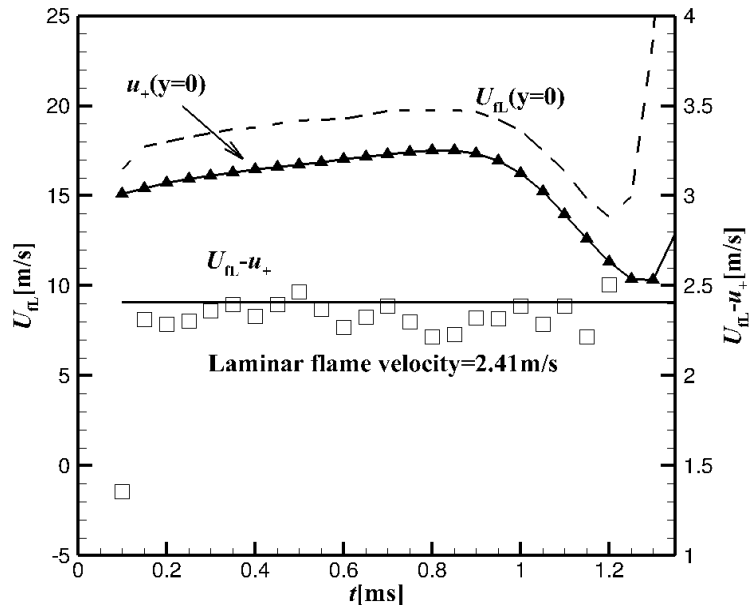

**Fig. B1.** The flame velocity at the tube axis $U_{fL}(y = 0)$ and the unburned gas velocity at 0.5 mm ahead of the flame $u_+(y = 0)$ for a flame initiated by planar ignition in a semi-open tube. The horizontal line shows the laminar flame velocity $U_f = 2.41 m/s$, the empty squares are $U_{fL}(y = 0) - u_+(y = 0)$.